\documentclass{article}  

\usepackage[margin=1in]{geometry}

\usepackage{graphicx} 
\usepackage{amsmath} 
\usepackage{amssymb}  
\usepackage{booktabs}
\usepackage[caption=false,font=footnotesize]{subfig}
\usepackage[linesnumbered,ruled,vlined]{algorithm2e}
\usepackage{xcolor}
\usepackage{multirow}
\usepackage{array}
\usepackage{authblk}

\graphicspath{{Figures/}}

\newcommand{\SO}[1]{\ensuremath{\textrm{SO}(#1)}}

\newcommand{\so}[1]{\ensuremath{\mathfrak{so}(#1)}}

\def\R{\ensuremath{\mathbb{R}}}

\def\vect{\ensuremath{\textrm{vec}}}
\def\trace{\ensuremath{\textrm{trace}}}

\begin{document}


\title{Machine learning based state observer for discrete time systems evolving on Lie groups }
\date{}
\author{Soham Shanbhag\thanks{sshanbhag@kaist.ac.kr} }
\author{Dong Eui Chang\thanks{Corresponding author, dechang@kaist.ac.kr}}
\affil{School of Electrical Engineering, Korea Advanced Institute of Science and Technology, Daejeon, Republic of Korea}

\maketitle

\begin{abstract}
    In this paper, a machine learning based observer for systems evolving on manifolds is designed such that the state of the observer is restricted to the Lie group on which the system evolves.
    Conventional techniques involving machine learning based observers on systems evolving on Lie groups involve designing charts for the Lie group, training a machine learning based observer for each chart, and switching between the trained models based on the state of the system.
    We propose a novel deep learning based technique whose predictions are restricted to a measure $0$ subset of Euclidean space without using charts.
    Using this network, we design an observer ensuring that the state of the observer is restricted to the Lie group, and predicting the state using only one trained algorithm.
    The deep learning network predicts an ``error term'' on the Lie algebra of the Lie group, uses the map from the Lie algebra to the group, and uses the group action and the present state to estimate the state at the next epoch.
    This model being purely data driven does not require the model of the system.
    The proposed algorithm provides a novel framework for constraining the output of machine learning networks to a measure $0$ subset of a Euclidean space without chart specific training and without requiring switching.
    We show the validity of this method using Monte Carlo simulations performed of the rigid body rotation and translation system.

    \textbf{Keywords:} Observer design, Machine learning, Lie groups
\end{abstract}


\section{Introduction}

State estimation is an active area of research due to its importance in feedback based control.
Many important systems evolve on Lie groups.
A common example of this is the state of a quadcopter, which lies on the special Euclidean group.
Estimation algorithms for systems evolving on Lie groups have to be designed such that the estimate of the state also evolves on the Lie group.

Classical state estimation techniques usually design a deterministic observer for the state of the system under consideration based on the system dynamics \cite{MahonHP2008,LagemTM2010,KhosrTML2015,VascoCSO2010,WangT2020a,BrasISSO2013}.
The design of such observers is based on some notion of error defined on the Lie group converging to $0$.
Due to the deterministic nature of these observers, they do not consider noise in the system and measurements.
Moreover, considering noise in the observer for systems evolving on Lie groups requires noise models on the group, which are more complicated than their Euclidean counterparts.
Alternatively, observers can also be designed such that the observer state does not lie on the Lie group, but converges asymptotically to it \cite{Chang2021,ParkPKC2021,ShanbC2022a}.
Such observers are designed based on the Kalman Filter and other noise rejecting filters for good performance in the presence of noise.
These estimators also allow a global region of convergence.

Another disadvantage of classical observer design techniques is that they require an accurate model for the system.
Modelling of such a system for discrete time systems involves first modelling a continuous time system using the Newton's laws of motion or Lagrangian mechanics.
The discrete time model is then generated by discretizing the continuous time system using a discretization scheme which respects the Lie group on which the system evolves.
Clearly, constructing the dynamical models for systems which are complex using classical techniques faces significant difficulty due to the complexity of the system.
Moreover, the discretization process also introduces errors in the modelling of the system.

Due to the advances in computing technologies, machine learning is a viable technique for tackling such problems.
Since models are not required for machine leraning based methods, they can design effective estimators for complex systems.
Moreover, since deep learning methods are regression based methods designed to not overfit on the data used for training, they inherently reject noise in the measurements.
However, the literature concerning use of machine learning based algorithms for state estimation is sparse.
The authors in \cite{DingYWWYN2011} assume linear system dynamics and design a data driven observer for this system.
The authors in \cite{RamosDMdB2020} use a deep learning based auto-encoder model for state observer, where the encoder generates a linear system which is a latent space representation of the system, and the decoder generates the estimates using the latent space representation of the estimator of the linear system. 
Similarly, the authors in \cite{PeralN2021} extend this using unsupervised learning based mapping to address the shortcomings of \cite{RamosDMdB2020}.
However, the estimates by the observers proposed in the literature are not restricted to the Lie groups on which the systems evolve.

In this paper, we propose a machine learning based method for state estimation which respects the Lie group.
We note that machine learning based methods are not able to predict state on measure 0 subsets of Euclidean spaces.
To handle such constrained sets, machine learning algorithms predict state in a lower dimensional set, and these estimates are mapped to the measure $0$ sets in the higher dimensional Euclidean space.
However, for many subsets of Euclidean spaces, this technique may require the use of multiple maps to cover the subset, and an algorithm trained for each map, along with a switching rule between the maps.
To remedy the problem of the estimates of the machine learning based method not restricted to the Lie group, we propose a network whose estimate lies on the Lie algebra of the Lie group on which the system evolves.
We use the previous estimate of the state, the exponential map from the Lie algebra to the Lie group and the group action of the Lie group to arrive at the next estimate of the state.
This technique is a novel technique which allows us to design observers without the use of multiple maps and multiple trained networks.
We validate the technique using Monte Carlo simulations of a rigid body system.
A benefit of this approach is that the estimator network has a ResNet like structure, which has shown to improve neural networks by allowing them to be trained easily and achieve higher accuracy \cite{HeZRS2016}.

The structure of the paper is as follows.
Section~\ref{sec_observer} introduces the system considered for state estimation.
The system considered in Section~\ref{sec_observer} is a general system evolving on a Lie group.
This section also presents the observer network for state estimation and loss function used for training the same.
Section~\ref{sec_SO3} shows an example of the proposed observer for the rigid body system dynamics evolving on the special Euclidean group.
We present our network and various maps required for designing the network in this section for the special Euclidean group.
We also present simulations with a measurement scheme.
The effect of noise on the estimates is also shown.
Finally, Section~\ref{sec_conc} concludes the paper.
We summarise the results here, list the benefits and drawbacks of the proposed observer, comment of the benefits afforded due to the exponential map, and present possible future topics of research related to this topic.


\section{Proposed machine learning based observer}\label{sec_observer}
\subsection{Preliminaries}
Consider the dynamical system evolving on a Lie group $G$ given by the difference equation
\begin{align}\label{sys_main_mani}
    x_{k+1} = f(x_k, u_k, \mu),\quad \quad y_k = h(x_k, \nu),
\end{align}
where $x_k \in G$ is the state of the system at epoch $k$, $u_k \in \R^m$ is the control input to the system at epoch $k$, $y_k \in \R^l$ is a measurement of the state of the system at epoch $k$, and $\mu$ and $\nu$ are noise representing process and measurement noise in the system.
A discrete time observer for this system is of the form
\begin{align*}
    \bar{x}_{k+1} = \bar{f}(\bar{x}_k, y_{k + 1}, u_k),
\end{align*}
where $\bar{x}_k \in G$ is the estimate of the state $x_k$ on $G$. For an ideal estimator, $\lim_{k \to \infty} E(\bar{x}_k, x_k) = 0$, where $E: G \times G \to \R$ is the error between two elements of $G$ defined on $G$.

\begin{figure}[t]
    \centering
    \resizebox{0.5\linewidth}{!}{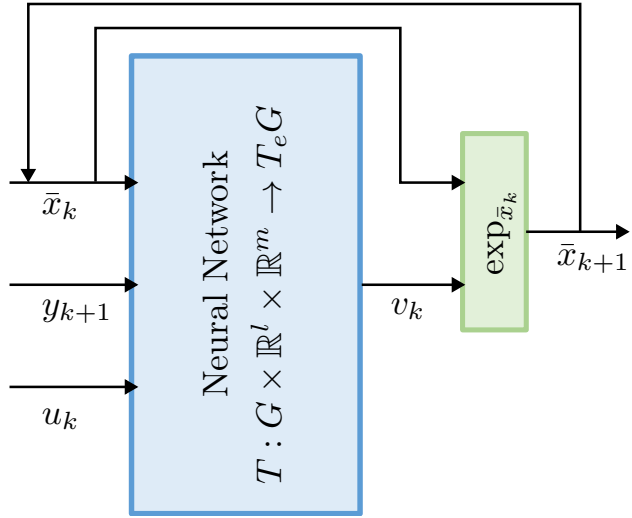}
    \caption{Proposed observer model for state estimation.}
    \label{fig_network}
\end{figure}

A machine learning based approach for state estimation is beneficial if we do not have information regarding the state update equation~\eqref{sys_main_mani}.
To design an observer for system~\eqref{sys_main_mani} using machine learning, we design a deep learning based network $T: G \times \R^l \times \R^m \to T_eG$, which is used to estimate the function $\bar{f}$ using properties of the Lie group, such that the estimated state is restricted to $G$.

\subsection{Observer design}
From the strong Whitney embedding theorem \cite{Whitn1944}, the Lie group $G$ can be smoothly embedded into a higher dimensional Euclidean space.
Denote the dimension of this ambient Euclidean space by $n$, and the embedding from $G$ to $\R^n$ by $\pi: G \hookrightarrow \R^n$.
Fig.~\ref{fig_network} shows the observer structure used for state estimation.
To estimate the state of the system at epoch $k + 1$, the input to the model is the state estimate $\pi(\bar{x}_k) \in \R^n$ at epoch $k$, the measurement $y_{k+1} \in \R^l$ at epoch $k + 1$, and the input to the system $u_k \in \R^m$ at epoch $k$.
The model is trained to output $v_k \in T_eG$, a ``residual error'', such that the estimate of the state at epoch $k + 1$ is arrived at by the equation
\begin{align}\label{eq_est_update}
    \bar{x}_{k+1} = \exp_{\bar{x}_k}(v_k) = \bar{x}_k \cdot \exp(v_k),
\end{align}
where $\cdot$ is the group action on $G$, and $\exp: T_eG \to G$ is the exponential map.

\begin{algorithm}[b]
    \SetKw{To}{ to }
    \SetKw{In}{in }
    \DontPrintSemicolon

    \KwData{$N$ sequences: $\{ \{u_{k-1}, y_{k}\}, k=1,\ldots,M \}$ and their corresponding ground truth states}
    Initialize parameters $\Theta$ of $T$ randomly \;
    \For{$iter \gets 1 \To iter^{max}$}
    { 
        \ForEach{Sequence}
        { 
            Set $\bar{x}_0 = e \in G$\;
            \For{$k \gets 0 \To M$}
            { 
                Compute $v_k = T(\bar{x}_k, y_{k+1}, u_k)$\;
                Compute $\bar{x}_{k+1} = \exp_{\bar{x}_k}(v_k)$\;
            }
        }
        Calculate $\mathcal{L}$ using \eqref{eq_training_loss}\;
        Update $\Theta$ using gradient descent to minimize $\mathcal{L}$\;
    }

    \caption{Learning $T$ from data.}
    \label{alg_learningT}
\end{algorithm}

Defining $\pi(\bar{x}_k)$ also allows the addition and subtraction operators to be defined, which may not be defined on the Lie group.
Using the embedding map, the loss between two elements of $G$ is defined as
\begin{align*}
    l(g_k, g'_k) = \| \pi(g_k) - \pi(g'_k) \|, \quad g_k, g'_k \in G,
\end{align*}
This loss function can be used for training of the function approximator.
Using this loss function, assuming $N$ sequences each of length $M$ available for training, the loss used for training is
\begin{align}\label{eq_training_loss}
    \mathcal{L} = \frac1{MN} \sum_{i=1}^{N} \sum_{k=1}^{M} l(\bar{x}^i_k, x^i_k)^2,
\end{align}
where the superscript denotes the sequence.
This loss is the mean squared error loss defined using the embedding map on $G$.

The procedure for designing a state observer using the proposed method is shown in Algorithm~\ref{alg_learningT}.
We denote the identity element of the group by $e$.


\section{Example: Rigid body system}\label{sec_SO3}

To show the estimation capability of the proposed observer, we show an example of an observer for the rigid body rotation and translation system.
Denote by $\SO{3}$ the set of orthogonal matrices of dimension 3 with determinant 1, and by $\so{3}$ its corresponding Lie algebra.
The elements of $\so{3}$ are skew-symmetric matrices in $\R^{3 \times 3}$.
The matrix representation of the cross product with a vector $v$ is denoted by $v_\times: \R^3 \to \so{3}$ such that for all $w \in \R^3$, $v \times w = v_\times w$.
The Euclidean norm of a matrix $A \in \R^{m \times n}$ is denoted by $\| A \| = \sqrt{\trace(A^T A)}$.
The vectorization of a matrix $A \in \R^{m \times n}$ is denoted by $\vect(A) \in \R^{m n}$.

\subsection{Preliminaries}

The state of a rigid body can be expressed using a rotational pose $R \in \SO{3} \subset \R^{3 \times 3}$ and a translational pose $p \in \R^3$.
We assume that the measurements of $R$ and $p$ are available as
\begin{align*}
    R^m = R \exp({\omega_1}_\times), \textrm{ and } p^m = p + \omega_2,
\end{align*}
where $\omega_1$ and $\omega_2$ are Gaussian noise with mean 0 and standard deviation $0.1$.
While we here assume the availability of $R$ and $p$ directly, the measurements in application depend on the measurement scheme, a commonly used example being the body frame measurement of at least three non-collinear markers whose inertial frame positions are known a priori in the case of camera or LiDAR sensors.
Since this would only affect the dimension of the input space to the model, we assume a more general measurement scheme to show applicability of the proposed observer.
We also assume availability of measurements of the angular velocity ($\Omega$) and the linear velocity ($v$) available as
\begin{align*}
    \Omega^m = \Omega + b^\Omega + \omega_3 \textrm{ and } v^m = v + b^v + \omega_4,
\end{align*}
where $\omega_3$ and $\omega_4$ are Gaussian noise with mean 0 and standard deviation $0.1$, and $b^\Omega$ and $b^v$ are measurement biases.
The biases $b^\Omega$ and $b^v$ are constant turn-on biases, and are a property of the operating conditions at the time of starting the sensors.
These measurement models are common measurement models of sensors like IMUs, velocity sensors, and accelerometers.

While the strong Whitney embedding theorem suggests that $\SO{3}$ can be embedded in a 6 dimensional Euclidean space, we here consider $\R^{9}$ as the ambient Euclidean space.
This allows us to define $\pi: \SO{3} \to \R^{9}$ trivially using the vectorization map on $\R^{3 \times 3}$, i.e. $\pi(R) = \vect(R)$, where $R$ is expressed as a matrix in $\R^{3 \times 3}$.
The state of the system is then $x = (\vect(R), p, b^\Omega, b^v) \in \R^{18}$.
Similarly, the measurements available for this system are $y = (\vect(R^m), p^m, \Omega^m, v^m) \in \R^{18}$.
Since $\Omega$ and $v$ are included in the measurements for this system, we consider no control input for the purpose of observer design.

In the chosen system, the state evolves on $\SO{3} \times \R^3 \times \R^3 \times \R^3$.
Since only the rotational state evolves on a manifold, we here only consider the exponential map for the rotational state.
The exponential map $\exp: \so{3} \to \SO{3}$ is defined as the matrix exponential.
The group action on $\SO{3}$ is defined as the matrix product.
Then, equation~\eqref{eq_est_update} can be written as
\begin{align*}
    \bar{R}_{k + 1} = \bar{R}_k \exp(\alpha_\times), \; \bar{p}_{k+1} = \bar{p} + \beta, \; \bar{b}_\Omega = \gamma, \;  \bar{b}_v = \delta,
\end{align*}
where the output of $T$ is $(\alpha, \beta, \gamma, \delta) \in \R^{12}$.
Assuming a sequence of length $M$ used for training, the loss function used for training is defined as
\begin{align*}
    \mathcal{L} &= \frac1{18 M} \sum_{k = 0}^M \| \bar{x}_k - x_k \|^2 = \frac1{18 M} \sum_{k = 0}^M \left(\| \bar{R}_k - R_k \|^2 \right.\\&\quad\quad\left.+ \| \bar{p}_k - p_k \|^2 + \| \bar{b}^\Omega_k - b^\Omega_k \|^2 + \| \bar{b}^v_k - b^v_k \|^2 \right).
\end{align*}

\subsection{Deep learning training}

To train the machine learning network, we design a network with structure as in Fig.~\ref{fig_network}.
Since the network would be repeated for each epoch during training and testing, using fully connected layers for $T$ leads to the vanishing/exploding gradients problem common in recurrent neural networks.
To mitigate this problem, we use a Gated Recurrent Unit (GRU) \cite{ChovGBBSB2014}.
However, the outputs of GRU are restricted to $[-1, 1]$ due to use of sigmoid and tanh functions in the GRU layers.
Hence, we use a fully connected layer as the final layer to remove this restriction.
We use two GRUs, each of hidden vector size 512, followed by a fully connected layer to output 12 dimensional vectors.
The network structure is shown in Table~\ref{tab_network}.

\begin{table}[t]
    \centering
    \begin{tabular}{lrr}
        \toprule
        Layer & Input size & Output size\\
        \midrule
        GRU & 36 & 512\\
        GRU & 512 & 512\\
        FC & 512 & 12\\
        \bottomrule
    \end{tabular}
    \caption{Network structure for observer for rigid body system. FC: Fully connected, GRU: Gated Recurrent Unit.}
    \label{tab_network}
\end{table}

\begin{figure}[!ht]
    \centering
    \subfloat[][Error in rotational state.]{\resizebox{0.45\linewidth}{!}{\input{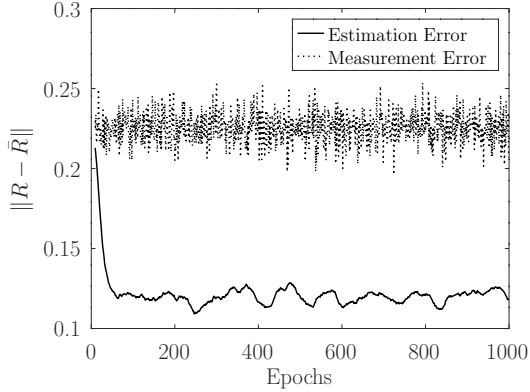}}}\quad
    \subfloat[][Error in position.]{\resizebox{0.45\linewidth}{!}{\input{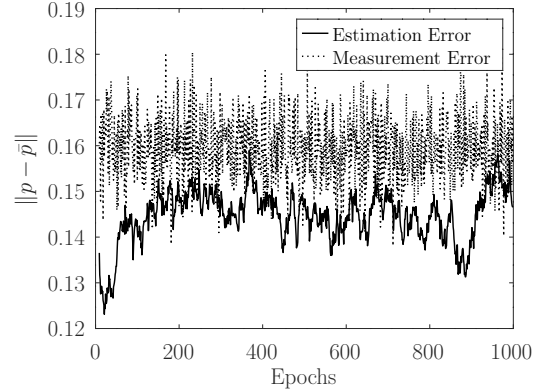}}}\\
    \subfloat[][Error in the estimate of bias in angular velocity.]{\resizebox{0.45\linewidth}{!}{\input{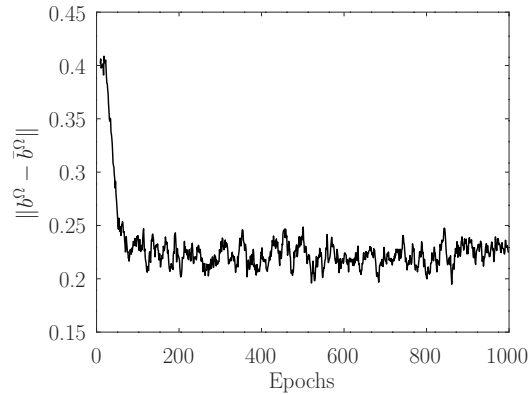}}}\quad
    \subfloat[][Error in the estimate of bias in linear velocity.]{\resizebox{0.45\linewidth}{!}{\input{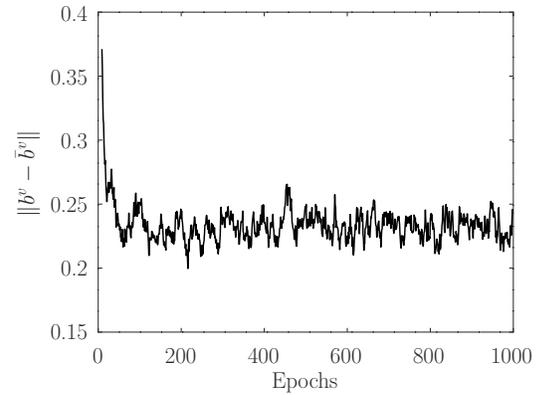}}}\\
    \subfloat[][Distance of the rotational state estimate from $\SO{3}$.\label{fig_dist_SO3}]{\resizebox{0.45\linewidth}{!}{\input{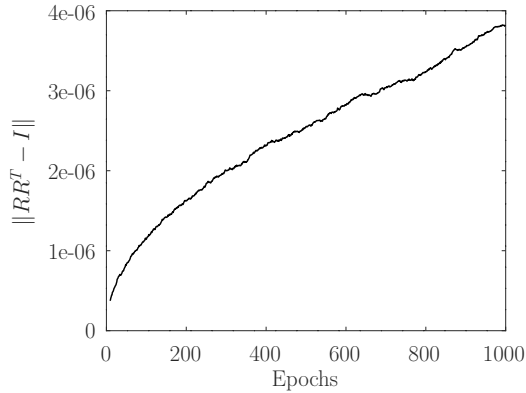}}}
    \caption{State estimation results for the machine learning based observer for the rigid body system.}
    \label{fig_results}
\end{figure}


The proposed observer is a machine learning based observer, hence requires data to be trained.
Common data generation methods for the rigid body system include experiments using a sensor like the Intel Realsense T265, using a simulator simulating rigid body dynamics like Gazebo, or using some other simulation algorithms.
We generate data using the system equations for a rigid body system.
The system equations are given by
\begin{align*}
    \dot{R} = R \Omega_\times, \quad \dot{p} = Rv,
\end{align*}
where $R \in \SO{3}$ is the rotational state of the body with respect to an inertial frame, $p \in \R^3$ is the translational state of the body with respect to the same inertial frame, $\Omega \in \R^3$ is the angular velocity of the body in the body frame, and $v \in \R^3$ is the linear velocity of the body in the body frame.

For every simulation sequence, we consider a random initial rotational position in $\SO{3}$ and translational position in $\R^3$.
We also sample biases from the uniform distribution $\mathcal{U}_{[-10, 10]}^3$.
At every epoch, we sample the angular velocity and the linear velocity from the uniform distribution $\mathcal{U}_{[-1, 1]}^3$.
We integrate the system from epoch $k$ to $k+1$ using the Explicit Runge-Kutta method of order 5(4) and generate 20000 sequences of length 100 each.
This dataset is used for training, testing and validation of the neural network.
These sequences do not contain any noise.
We split the dataset into three parts of size 80\%-10\%-10\% of the complete dataset and use the 80\% part for training, 10\% part for validation, and the rest for testing.
We similarly generate 1000 inference sequences with each sequence of length 1000 with the measurements corrupted by noise of mean $0$ and standard deviation $0.1$.
To train the network, we use the AdamW optimizer \cite{LoshcH2018} with learning rate $3 \times 10^{-4}$ and weight decay $0.1$.
At every training epoch, we calculate the error on the validation set and use the model having the smallest validation error as the final model.

\subsection{Results}

We plot the mean of the errors of the estimated state of the 1000 sequences in Fig.~\ref{fig_results}.
For the purpose of plotting, the first 10 epochs are skipped to allow the initial stabilization of the observer.
We see that the observer estimates the bias in the system measurements to an accuracy of approximately $0.2$ when the noise in the measurements is $0.1$.
Moreover, while the training set of the observer contains no noise, the observer is able to estimate the state correctly in the presence of noisy measurements of the state.
We also note that the observer improves the estimate of the state as compared to the measurements by a significant amount. 
The estimates show a 47.83\% reduction in mean error as compared to the measurements in case of the rotational pose and a 6.25\% reduction in mean error in case of position.

We also show that the proposed observer respects the manifold on which the system evolves.
The system here evolves on $\SO{3} \times \R^3 \times \R^3 \times \R^3$, hence showing that the rotational state is restricted to $\SO{3}$ suffices.
We plot the distance of the observer from the set $\SO{3}$ in Fig.~\ref{fig_dist_SO3}.
This error is defined as $\| R R^T - I \|$, which is $0$ for elements of $\SO{3}$.
It can be seen that the distance of the estimate of the rotational state from the manifold is of the order of $10^{-6}$, which is due to the precision of the programming language used.
An error of $10^{-6}$ is negligible for application purposes.

To show the effect of noise on the observer, we also perform simulations varying the noise added to the measurements.
This noise is modelled as a Gaussian noise with mean $0$ and standard deviation $\sigma$.
The mean error in the measurements and the estimates from the true value is shown in Table~\ref{tab_noise_varying}.
We see that the state estimates reduce the noise in the measurement by 50\% -- 60\% even in presence of high amount of noise.
However, the presence of noise does affect the estimates of the states, which is expected.
Moreover, from Table~\ref{tab_noise_varying}, it can be seen that the distance between the state and the manifold is of the order $10^{-6}$ irrespective of the noise.
This is a significant improvement over existing machine learning based algorithms, which are unable to restrict the estimated state to a Lie group, or require multiple charts and switching logic.

\begin{table*}[t]
    \centering
    \begin{tabular}{*{10}r}
        \toprule
        \multirow{2}{*}{$\sigma$}& \multicolumn{6}{c}{Error: $\| x - x_{true}\|$} & \multicolumn{2}{c}{\% error reduction} & \multirow{2}{*}{ $\| R R^T - I \|$}\\
        \cmidrule(l){2-7}
        \cmidrule(l){8-9}
        & $R^m$ & $\bar{R}$ & $p^m$ & $\bar{p}$ & $\bar{b}_\Omega$ & $\bar{b}_v$ & $R^m$ vs $\bar{R}$ & $p^m$ vs $\bar{p}$ &\\
        \midrule
        0.1 & 0.23 & 0.12 & 0.16 & 0.15 & 0.22 & 0.23 & 47.83 & 6.25 & $4 \times 10^{-6}$\\
        0.2 & 0.45 & 0.18 & 0.32 & 0.18 & 0.37 & 0.33 & 60.00 & 43.75 & $4 \times 10^{-6}$\\
        0.3 & 0.66 & 0.22 & 0.48 & 0.20 & 0.51 & 0.46 & 66.67 & 58.33 & $4 \times 10^{-6}$\\
        0.4 & 0.88 & 0.36 & 0.64 & 0.26 & 0.63 & 0.59 & 59.09 & 59.38 & $4 \times 10^{-6}$\\
        0.5 & 1.08 & 0.54 & 0.80 & 0.34 & 0.74 & 0.74 & 50.00 & 57.50 & $3 \times 10^{-6}$\\
        \bottomrule
    \end{tabular}
    \caption{Mean error in simulations with varying noise.}
    \label{tab_noise_varying}
\end{table*}


\section{Discussions and Conclusion}\label{sec_conc}

Due to the $\exp$ map, the state of our observer is restricted to the manifold.
The $\exp$ map also performs similar to skip connections used in ResNets, which have shown to improve neural networks by allowing them to be trained easily and achieve higher accuracy.
Due to the network being trained using deep learning, the network allows the design of an observer without knowledge of the system equations.
The design of the network only requires knowledge of the Lie group to design the $\exp$ map.
We also note that due to the inherent ability of neural networks to reject noise, the network does not require information of the noise in the system and the measurements during training.
It can estimate the state of the system even in the presence of noise in the measurements.

Another similar approach to observer design on Lie groups involves using local charts of the group.
The estimate of the state can be estimated in the local charts, and then the corresponding state on the group can be arrived at using the chart map.
However, this method requires a different network to be trained for each chart, which is cumbersome.
The proposed $\exp$ map based observer allows us to mitigate this problem of multiple charts due to embedding the manifold in the Euclidean space, while respecting the manifold.

However, due to being a deep learning based method, the training of this algorithm requires a large amount of data.
This data can be collected through experiments or simulations.
Also, while the estimation algorithm using a well-trained network is globally convergent, no guarantees can be provided about the rate of convergence of the network.

We have shown the estimation properties of the network with an example, where the network is correctly able to estimate the state and biases in the system.
While we have assumed the measurements of the states $(R, p)$ directly available here, the network can be easily modified for a different measurement scheme using transfer learning, using a preprocessing step if the measurement equations are known, or a trained network to generate $R$ and $p$ from measurements used as a preprocessing step.

Possible future work in this problem may include reducing the amount of data needed.
Also, the properties of convergence of the network should be studied.
Another possible extension of the above result is to design constrained deep learning networks.
Since manifolds are level sets of continuous differentiable functions, the above result can be extended to constrained subsets of the Euclidean space in a straightforward manner.

\bibliographystyle{plain}
\bibliography{References}

\end{document}